\documentclass[prc,showpacs,twocolumn,aps]{revtex4}

\usepackage{bm}
\usepackage{amsmath}
\usepackage{epsfig}
\usepackage{psfig}
\usepackage{graphics}
\usepackage{graphicx}

\begin{document}

\preprint{TRI-PP-04-15}


\title{Effects of Rescattering in $(e,e'p)$ Reactions within a Semiclassical Model}

\author{C.~Barbieri}
  \email{barbieri@triumf.ca}
  \homepage{http://www.triumf.ca/people/barbieri}
  \affiliation
     {TRIUMF, 4004 Wesbrook Mall, Vancouver, 
          British Columbia, Canada V6T 2A3 \\  }

\author{L.~Lapik\'{a}s}
  \affiliation
     {NIKHEF, P.O. Box 41882, 1009 DB Amsterdam, The Netherlands \\  }

\date{\today}

\begin{abstract}
The contribution of rescattering to  final state interactions
in $(e,e'p)$ cross sections is studied for medium and high missing
energies using a semiclassical model.
This approach considers two-step processes
that lead to the emission of both nucleons.
 The effects of nuclear transparency are accounted for in a Glauber
 inspired approach and
the dispersion effects of the medium at low energies are included. 
 It is found that rescattering is strongly reduced in parallel
kinematics. 
 At high missing energies and momenta, the distortion of the short-range
correlated tail of the spectral function is dominated by a rearrangement
of that strength itself.
In perpendicular kinematics, a further enhancement of the experimental
yield is due to strength that is originally in the mean field region.
 This contribution becomes negligible at large missing momenta.
\end{abstract}

\maketitle

\section{Introduction}

 Nuclear correlations strongly influence the dynamics
of nuclear systems~\cite{PandSick97,DB04}.
 In particular, the repulsive core at small 
internucleon distances has the effect of depleting
the shell model orbitals and inducing high momentum
components in the nuclear wave functions~\cite{WHA,PieperO16}.
 The main effects of short-range correlations (SRC) consist
in shifting a sizable amount of spectral strength, 
about 10-15\%~\cite{bv91}, to very high missing energies and momenta,
together with increasing the binding energy~\cite{Wim03}. 
The resulting reduction of the occupation numbers of the deeply bound
orbitals appears to be fairly independent of the given subshell
and of the size of the nucleus, except for a slight increase with
the central density of the system.
 Theoretical studies of the distribution of short-range correlated
nucleons for finite nuclei have been carried out using a local density
approximation (LDA) by Benhar et al.~\cite{Benhar} and with
many-body Green's functions by M\"uther et al.~\cite{WHA}.
  These calculations suggest that most
of the missing strength is found along a ridge in the momentum-energy plane
($p_m$-$E_m$) which spans several hundreds of MeV/c (and MeV).
Such behavior is confirmed by recent experimental data~\cite{Frick04}.

It is important to note that the depletion of spectroscopic factors for
closed shell orbitals observed near the Fermi energy is more substantial than
the 15\% reduction discussed above~\cite{Louk93} due to long range effects
like the coupling to collective modes~\cite{DB04}.
 We note that this reduction also tends to be weaker for loosely bound
orbitals like halo states~\cite{FaddO16,greg04}.
As a consequence, it becomes particularly interesting to study the
spectral distribution in heavy nuclei where
the mean field single particle orbitals extend to regions far from
the Fermi level
and tend to decouple from surface effects.
A measurement of the spectral function for the complete mean field
region of ${}^{208}$Pb has been
undertaken recently at NIKHEF~\cite{Marcel}. 
In this region the single particle orbitals are sensibly fragmented and
it is required to probe missing energies up to 100~MeV~\cite{Marcel} or more.
 Further detailed information on SRC could 
be obtained from $(e,e'p)$ experiments
that directly search for the missing strength at very high missing momenta
and energies.
This is particularly appealing since the details of its distribution
strongly influence the binding energy of finite nuclei and nuclear
matter~\cite{Wim03}.
 Besides, it could shed more light on how much the interior of large
nuclei is sensitive to the effects of finite size and
proton-neutron asymmetry.

\begin{figure}[t]
  \begin{center}
    \includegraphics[height=0.24\textheight]{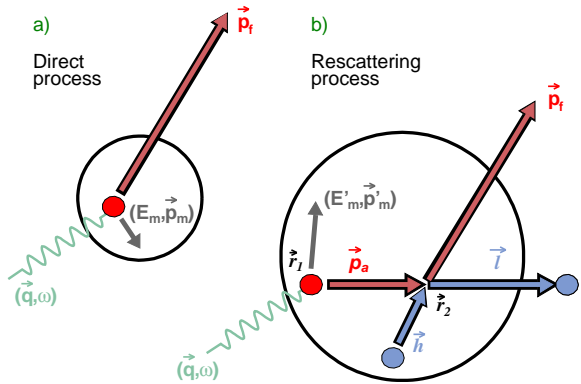}
    \caption{ \label{fig:TotRes}
       (Color online) 
       Schematic representation of the direct knockout of a proton (a),
      given by the PWIA, and the contribution from a two-step
      rescattering (b). In the latter a proton or neutron is emitted
      with momentum ${\bf p}_a$ and different missing energy and momentum
      $(E_m',{\bf p}_m')$. Due to a successive collision, a proton
      is eventually detected with the same momentum ${\bf p}_f$ seen in the
      direct process.
  }
    \end{center}
\end{figure}

Unfortunately, past measurements of the short-range correlated tail
by means of $(e,e'p)$ reactions have been limited due to the enormous
background that is generated  by final state interactions (FSI), see for
example Refs.~\cite{MIT-BATES1,MIT-BATES2}. 
The issue of how to minimize the FSI has been recently addressed
in Ref.~\cite{E97proposal}. There, it was suggested that FSI
in exclusive $(e,e'p)$
cross sections are dominated by two-step rescattering processes like the one
depicted in Fig.~\ref{fig:TotRes}. This becomes particularly 
relevant when regions of small spectral strength are probed
in perpendicular kinematics%
~\footnote{In this work we refer to `parallel' and `perpendicular' kinematics
in terms of the angle between the momentum of the virtual photon ${\bf q}$
and the missing momentum ${\bf p}_m$ (as opposed to the outgoing
proton ${\bf p}_f$).
}.
%
A study of several kinematic conditions shows that 
the rescattered nucleons can move spectral strength in the
$p_m$-$E_m$ plane, from the top of the ridge toward regions where the
correlated strength is small, therefore submerging the direct signal
in a large background noise.
Other possible contributions that
involve the excitation of a $\Delta$ resonance
are expected to be more sensitive to transverse degrees of freedom.
Parallel kinematics tend to be more clean due to the high momentum
that is required for the detected proton.
 In Ref.~\cite{E97proposal}, it was suggested that the contribution
of rescattering can be diminished by choosing parallel kinematics
and taking advantage of modern electron beam facilities.
 New data were subsequently taken
in these conditions by the \hbox{E97-006} collaboration at Jefferson
Lab~\cite{danielaElbaDiss,danielaPRL,Frick04} for a set of nuclei ranging
from carbon to gold.
Clearly, FSI could still play a role even in parallel kinematics
and need to be properly addressed before the relevant physical information
is extracted from the experiment.
We note that a similar dependency of the FSI on the kinematics
is also predicted in Ref.~\cite{MuetherFSI}
for $(e,e'NN)$ reactions in superparallel kinematics.

 The issue of computing the effects of rescattering
has been considered recently by means of the
multistep dynamics approach~\cite{VivianPV} and by using Glauber
theory~\cite{NicolaevGL,Ciofi4He,Ryckebusch03}.
  All these calculations suggest that multiple rescattering contributions
(more than two-steps) are relatively small in light nuclei like
${}^{12}$C but can become relevant for large systems.
Interference effects between FSI and SRC correlations
can also play a role~\cite{Ciofi4He}.
 However, all these effects were seen to be reduced in parallel kinematics.
In Ref.~\cite{Marcel}, the scattered proton was detected at energies
at which a full distorted wave calculation, in terms of an optical
potential, is required.  However, rescattering processes leading to
the emission of two nucleons (one of which is not detected) can 
lead to the reappearance of a part of the experimental strength
absorbed by inelastic processes.
 This effect was investigated in terms of a semiclassical model inspired
by the work of Ref.~\cite{VjPi}.
Even if very different kinematical situations were considered,
the reaction mechanism included in the latter approach is
the same as pointed out in Ref.~\cite{E97proposal} and Fig.~\ref{fig:TotRes}.
Some partial simplifications occur for high energy protons,
since the relevant effects of the medium are limited to Pauli blocking.
Therefore this approach offers a valid starting point to investigate the FSI
effects needed in the analysis of data at large missing energy 
and momenta.
Other effects such as meson exchange currents and the excitation
of resonances also need to be investigated. However, these are beyond
the scope of the present paper and will be considered in future work.
In this paper, we consider the approach of Ref.~\cite{Marcel}
and extend it to high missing energies.
 We then apply it to the kinematics of both the NIKHEF~\cite{Marcel} ad the
E97-006~\cite{danielaElbaDiss,danielaPRL} experiments to evaluate the
importance of two-step processes for the different kinematics employed.

The model for computing the contribution of rescattering is
depicted in Sec.~\ref{rescatt_formula}, together with a discussion of 
the inclusion of the absorption effects in terms of nuclear transparency. 
 A practical application requires the knowledge of the in-medium
differential cross section, which is calculated in Sec.~\ref{sec:in-med_xsec}
by extending the approach of Ref.~\cite{VjPi}.
 Secs.~\ref{sec:respb208} and~\ref{sec:resE97-006} report on the
results for the kinematics used in the above experiments, at medium 
and high missing energies respectively.
Conclusions are drawn in Sec.~\ref{sec:concl}.


\section{Model}
\label{rescatt_formula}

This work considers contributions to the experimental yield that
come from two-step mechanisms
 in which a reaction $(e,e'a)$ is followed by a scattering process
from a nucleon in the medium, $N'(a,p)N''$, eventually leading
to the
emission of the detected proton. In general, $a$ may represent
a nucleon or another possible intermediate particle.
In this work we will only consider the channels in which $a$
is either a proton (with $N'=p$ or $n$) or a neutron.
 In the following we will also use the letter $a$ to label
the possible open channels.

Following the semiclassical approach of Refs.~\cite{VjPi,Marcel},
the contribution to the cross section coming from rescattering
through the above channels is written as
\begin{widetext}
\begin{eqnarray}
    { d^6 ~ \sigma_{rescatt.}
     \over
     dE_0 \; d\Omega_{\hat{k}_o}  dE_f \; d \Omega_{\hat{p}_f} } &=&  
    \sum_a
    \int d {\bf r}_1 \int d {\bf r}_2 \int_{0}^{\omega} d T_a \; \;
  \rho_N({\bf r}_1) \; 
    { K \; S^h_a(p'_m,E'_m) \; \sigma^{cc1}_{ea}
    \over
    M \; ( {\bf r}_1 - {\bf r}_2 )^2 } 
    g_{aN'}(|{\bf r}_1 - {\bf r}_2|) \; \; \;
\nonumber  \\
 &\;& \times 
 P_T(p_a; {\bf r}_1 , {\bf r}_2 )
\rho_{N'}({\bf r}_2) \; 
    { d^3 ~ \sigma_{a N'}
    \over
    dE_f \; d \Omega_{\hat{p}_f} } \;
P_T(p_f; {\bf r}_2 , \infty) \; ,
\label{eq:TotRes}
\end{eqnarray}
\end{widetext}
where $(E_o,{\bf k}_o)$ and  $(E_f,{\bf p}_f)$ represent the four-momenta
of the detected electron and proton,  respectively.
Eq.~(\ref{eq:TotRes}) assumes that the intermediate particle $a$ is
generated in plane wave impulse approximation (PWIA)
by the electromagnetic current at a point ${\bf r}_1$
inside the nucleus. Here $K=|{\bf p}_a|E_a$ is a phase space factor,
$S^h_a(p'_m,E'_m)/M$ is
the spectral function of the hit particle $a$ normalized to one [i.e.,
$M=N$($Z$) if $a$ is a neutron (proton)]
and $\sigma^{cc1}_{ea}$
the off shell electron-nucleon cross section, for which
we have used the $cc1$ prescription of de~Forest~\cite{deForest}.
The pair distribution functions
$g_{aN'}(|{\bf r}_1 - {\bf r}_2|)$ account for the joint probability of
finding a nucleon N' in ${\bf r}_2$ after the particle $a$ has been struck
at ${\bf r}_1$~\cite{gNN}.
The integration over the kinetic energy $T_a$ of the intermediate particle $a$
ranges from 0 to the energy $\omega$ transfered by the electron.
The nuclear transparency factor $P_T(p; {\bf r}_1 , {\bf r}_2)$ gives
the transmission
probability that the struck particle $a$ propagates, without any
interactions,
to a second point ${\bf r}_2$, where it scatters from the nucleon $N'$ with
cross section $d^3 ~ \sigma_{a N'}$. The whole process is depicted
in Fig.~\ref{fig:TotRes}b.
The point nucleon densities $\rho_N({\bf r})$ are normalized to either the 
number of neutrons or protons.
 These were derived from experimental charge distributions by unfolding
the proton size~\cite{density_tables}. 
We employed equal distributions for neutrons and protons, which is a
sufficiently accurate approximation since even in $^{208}$Pb neutron
and proton radii differ by less than 4\%.
%

 The nuclear transparency for the ejected nucleon was considered in
Ref.~\cite{VjPi} according to Glauber theory.  The
probability $P_T$ that a proton struck at ${\bf r}_1$ will travel
with momentum ${\bf p}$ to the point ${\bf r}_2$ without being
rescattered is given by
\begin{eqnarray}
\lefteqn{
   P_T(p;{\bf r}_1,{\bf r}_2) = 
   } & &
\nonumber \\
 exp \left\{ - \right. \int_{z_1}^{z_2} dz & &
    \left[ g_{pp}(|{\bf r}_1 - {\bf r}|) \; \tilde{\sigma}_{pp}(p,\rho({\bf r}))
           \; \rho_p({\bf r})   \right.
\label{eq:PT} \\
   & &+ \left.   g_{pn}(|{\bf r}_1 - {\bf r}|) \; \tilde{\sigma}_{pn}(p,\rho({\bf r}))
           \; \rho_n({\bf r}) \right]  \left. \right\} \; ,
\nonumber
\end{eqnarray}
where the z axis is chosen along the direction of propagation
$\hat{\bf p}$, an impact
parameter ${\bf b}$ is defined so that ${\bf r} = {\bf b} + z \hat{\bf p}$,
and $z_1$ ($z_2$) refer to the initial (final) position.
Eq.~(\ref{eq:PT}) differs from the standard Glauber theory
by the inclusion of the pair distribution functions
$g_{pN}(|{\bf r}_1 - {\bf r}|)$.
 In principle, the $g_{pN}$ functions should depend on the
density and on the direction of the inter-particle distance.
 However, these
effects has been shown to be negligible in Ref.~\cite{VjPi}.
In the present application we find that a simple two-gaussian
parameterization of the $g_{pN}$ can adequately fit the curves
reported there 
for nuclear matter at saturation density. 
The in medium total cross sections $\tilde{\sigma}_{pp}(p,\rho)$ and
$\tilde{\sigma}_{pn}(p,\rho)$ used in this work have been computed according
to Ref.~\cite{VjPi} 
and account for the effects of Pauli blocking,
Fermi spreading and the effective mass generated by the nuclear mean field.
 For energies above 300~MeV they have been extended to incorporate the
effects of pion emission~\cite{Pieperpriv}.

The nuclear transparency is defined 
as the average over the nucleus of the probability that the struck proton
emerges from the nucleus without any collision. This is related
to $P_T$ by
\begin{equation}
 T = {1 \over Z} \int d{\bf r} \rho_p({\bf r}) P_T(p;{\bf r},\infty) \; .
 \label{eq:T}
\end{equation}
It should be mentioned that the practical experimental definition of nuclear
transparency depends on the specific kinematics employed and that
Eqs.~(\ref{eq:PT}) and~(\ref{eq:T}) rigorously apply
only to the parallel case~\cite{Kelly86}.
 For the present approach, these are the right quantities to be included
in Eq.~(\ref{eq:TotRes}) since they describe the loss of flux in the
direction of propagatiom. 
 In the case of ${}^{208}{\rm Pb}$ and an outgoing proton with
energy $E_f \sim$~1.1~GeV (kinetic energy of $\sim$161~MeV),
Eq.~(\ref{eq:T}) gives $T =$~0.37.
 With $E_f \sim$~1.8~GeV, which is of interest for the calculations
of Sec.~\ref{sec:resE97-006},
$T =$~0.63 for ${}^{12}{\rm C}$ and $T =$~0.29 for ${}^{197}{\rm Au}$.
The PWIA contribution of the direct process, Fig.~\ref{fig:TotRes}a,
also needs to be corrected  for the nuclear transparency effects,
\begin{equation}
   { d^6 ~ \sigma_{PWIA(T)}
     \over
     dE_0 \; d\Omega_{\hat{k}_o}  dE_f \; d \Omega_{\hat{p}_f} }  
  ~=~  K \; \sigma^{cc1}_{ep} \; S^h_p(p'_m,E'_m) \; T   \; ,
\label{eq:PWIA}
\end{equation}
with $T$ given by Eq.~(\ref{eq:T}).
 Eq.~(\ref{eq:PWIA}) is consistent with the assumptions of
Eq.~(\ref{eq:TotRes}), from which it would be obtained if the
rescattering event was substituted with the limit of ${\bf r}_2$
to infinity.

\section{Evaluation of the in-medium nucleon-nucleon rate}
\label{sec:in-med_xsec}

 The last ingredient required
in Eq.~(\ref{eq:TotRes}) is the in medium cross section for the
rescattering  process, in which the particle $a$
(either a proton or a neutron) hits against a bound
nucleon in its way out, eventually leading to the emission of the
detected proton.
 This can be evaluated by extending the approach of Ref.~\cite{VjPi} to
describe the angular dependence of the ejected proton.
We start form the NN differential elastic cross section in free space
\begin{equation}
 { d  \sigma_{pN}  \over d \Omega_{\hat{p}_f} }
    = \left| f_{pN}(cos \; \theta) \right|^2  = 
  {  \left| \bar{\cal M}(s,t,u) \right|^2 \over  64 \; \pi^2 \; s  }
    \; ,
 \label{eq:free-dSig}
\end{equation}
where $s$, $t$ and $u$ are the Mandelstam invariants
($\sqrt{s}$ being energy in the c.m. system) 
and $\left| \bar{\cal M}(s,t,u) \right|^2$ represents the square of
the Lorentz invariant amplitude~\cite{PeSch} averaged (summed) over
the initial (final) spins. 
 For nucleon momenta above 1~GeV/c and for small angles, the scattering
amplitude in Eq.~(\ref{eq:free-dSig}) is well approximated by its
central part and can be written as
\begin{equation}
 f_{pN} = \frac{ p_f  \sigma^{tot}_{pN} }{4 \pi} (\epsilon_{pN} + i)
    exp\{-\beta^2_{pN} ({\bf p}_i - {\bf p}_f)^2 /2 \}
    \; ,
 \label{eq:fGlaub}
\end{equation}
where ${\bf p}_i$ and ${\bf p}_f$ are the initial and final momentum
of the scattered nucleon,
$\epsilon_{pN}$ is the ratio of the real to imaginary part of
the scattering amplitude
  and $\sigma^{tot}_{pN}$ is the total scattering
cross section.
At low energy the values of ${\cal M}(s,t,u)$ where extracted from
the SAID phase shift data analysis~\cite{SAID}.
 For the
pp case, we chose to keep the differential cross section constant
for angles smaller than 5${}^o$ and larger than 175${}^o$,
in order to avoid the Coulomb peak in the forward and backward directions.
 However, the results of
the present work are largely insensitive to this choice of the cut off
angle due to Pauli exclusion.
 The solutions of the SAID program were used for energies
in the laboratory system up to 1.6~GeV for pp scattering and 1.2~GeV
for the pn case, which are well contained within the range of validity
of this data base.
 At higher energies Eq.~(\ref{eq:fGlaub}) was used with parameters
$\sigma^{tot}_{pp}=$44.0~mb, $\sigma^{tot}_{pn}=$41.1~mb
and $\epsilon_{pp}=\epsilon_{pn}=$-0.48. The slope coefficients
were chosen by requiring that Eq.~(\ref{eq:free-dSig})
yields the correct values $\sigma^{el}_{pN}$ for the total elastic
cross section. This implies
\begin{equation}
 \beta^2_{pN} \simeq \frac{(1 + \epsilon_{pN}^2) \; {\sigma^{tot}_{pN}}^2}
           {16 \; \pi \; \sigma^{el}_{pN} }
    \; ,
 \label{eq:betaGlaub}
\end{equation}
where $\sigma^{el}_{pp}$ and $\sigma^{el}_{pn}$ were extracted from the
experiment~\cite{PDG}.  A direct comparison
for energies above 1~GeV showed that Eq.~(\ref{eq:fGlaub}) appropriately
approximates the data from the SAID data base, thus there exists an overlap
region where these approaches are both accurate and join smoothly.

In deriving the rescattering rate we assume 
that the interaction between the two nucleons
is localized enough so that the amplitude ${\cal M}(s,t,u)$ is
not altered by the presence of the surrounding nucleons.
 However, the medium can sensibly modify the cross section due to the
spectral distribution of the momentum of the hit nucleon and due to the
effects of Pauli blocking.
In the spirit of the local density approximation (LDA), which
underlies Eq.~(\ref{eq:TotRes}), the  momentum of
the hit nucleon, $N'$, is taken to be locally distributed as in infinite
nuclear matter. The density of the latter being the one of the point
${\bf r}_2$ where the collision occurs,
$\rho_{NM}=\rho_{N'}({\bf r}_2)$.
The effects of the nuclear surface are eventually included by integrating
over $\rho_{N'}({\bf r}_2)d{\bf r}_2$ in Eq.~(\ref{eq:TotRes}).
For the present purposes it is appropriate to further approximate
the symmetric nuclear matter with a free Fermi gas~%
\footnote{Note that in correlated nuclear matter a sizable number of
protons ($\sim$15\%) have momentum larger than $k_F$. However, the error
on the spreading effects generated by considering all of them inside
the Fermi sea did not appear to be relevant in Ref.~\cite{Marcel}.
This discrepancy also tends to reduce further for large nucleon energies.}.
The assumption of a completely filled Fermi sea is also consistent with
the Dirac-Brueckner-Hartree-Fock (DBHF) employed below.
Initially, the hit nucleon $N'$ is in the Fermi sea and therefore must have
a momentum ${\bf h}$ smaller than $k_F = (3\pi^2\rho_{NM}/2)^{1/3}$.
 At the same time the Pauli principle requires that the
particles in the final state will have momenta ${\bf p}_f$ and ${\bf l}$,
both larger than $k_F$.
 Among all the nucleons involved in the process,
${\bf p}_f$ refers to the detected proton while the others can be
either neutrons or protons depending on the channel $a$.
The probability per unit time of an event leading to the emission
of a proton with momentum ${\bf p}_f$ is obtained by imposing the
Pauli constraints and integrating over the unobserved
momenta ${\bf h}$ and ${\bf l}$,
\begin{widetext}
\begin{eqnarray}
\lefteqn{ 
   {d^3 P_{aN'} 
      \over dp_f \; d\Omega_{\hat{p}_f} } ~=~
  2 \; \theta(p_f - k_F)  \; L^3
      \int       \int       {d{\bf h} \;  d{\bf l} \over (2\pi)^6} 
      \; \theta(k_F - h) \theta(l - k_F) \;
         W_I
      }   & &
\nonumber \\
 &=& 2 \; p_f^2 \; \theta(p_f - k_F) 
   \int         {d{\bf h} \over (2\pi)^3} 
    \theta(k_F - h) \theta(l - k_F)
 \left. 
   { 1 \over  64 \pi^2}
   { \left| \bar{\cal M}(s,t,u) \right|^2
   \over
   E_a(p_a) E_{N'}(h) E_f(p_f) E_{N''}(l)}
   \delta(E_a + E_{N'} - E_f - E_{N''})
 \right|_{{\bf l} = {\bf p}_a + {\bf h} - {\bf p}_f} \; ,
\label{eq:P_dpf} 
\end{eqnarray}
\end{widetext}
where $L^3$ is the volume of a normalization box and,
for a free nucleon, $E_N(p)=(p^2 + m_N^2)^{1/2}$.
%
%
In Eq.~(\ref{eq:P_dpf}), $W_I$ is the probability per unit time for
the event $p_a^\mu + h^\mu \rightarrow p_f^\mu + l^\mu$,
which can be expressed in terms of ${\cal M}(s,t,u)$.
The inverse life time of the nucleon $a$ for energies below
the pion production threshold is related to Eq.~(\ref{eq:P_dpf}) by
\begin{equation}
 \frac{1}{\tau_a} = \sum_{N'=p,n} ~ \int d\Omega_{\hat{p}_f} \int d p_f 
{d^3P_{aN'} \over dp_f \; d\Omega_{\hat{p}_f} }  \; .
\end{equation}

 At low energies, a nucleon traveling through the medium acquires an
effective mass due to dispersion effects.
For infinite matter, this can be described
by a scalar field $U_S$ and the time component of a vector
field $U_V$~\cite{waleckaANP}. This particular approach allows to maintain
the relativistic framework adopted in Eq.~(\ref{eq:P_dpf}).
The values of $U_S$ and $U_V$ in nuclear matter were computed
in Ref.~\cite{MachUsVs} by solving
the DBHF equations.
The results were found to be consistent with the value of the non relativistic
effective mass extracted from nucleon-nucleus scattering.
At the energies considered by the DBHF calculations,
$U_S$ and $U_V$ are predicted to be essentially momentum independent.
Thus, the energy of a nucleon moving with momentum $p$ is given by
\begin{eqnarray}
  m_D(\rho_{NM}) &=& m_N + U_S(\rho_{NM})    \;  ,
\label{eq:Mdirac} 
 \\
  E_N(p,\rho_{NM}) &=& \sqrt{p^2 + m_D^2(\rho_{NM})} + U_V(\rho_{NM})  \; ,
\label{eq:Edirac} 
\end{eqnarray}
where $m_D$ is the Dirac effective mass.
 It should be noted that $U_S$ and $U_V$ have large and opposite
values. The success of the DBHF approach rely on a subtle cancellation
of their effects and require a self-consistent calculation of the
interaction with the medium~\cite{MachUsVs}.
As a consequence, the use of Eqs.(\ref{eq:Mdirac}) and~(\ref{eq:Edirac})
to compute the in medium cross section at low energy does
not guarantee a priori accurate predictions.
 In a relativistic model, this would
require a more elaborate calculation of the scattering amplitude
${\cal M}$ (see for example Refs.~\cite{MachLi,Fuchs01}).
 In the present work, we are simply interested in giving an approximate
treatment of the dispersion effects on the rescattering
of protons for the kinematics that will be considered
in Sec.~\ref{sec:respb208}.
 The kinematics relevant for studying the short-range
correlated tail involve much higher energies, where the nuclear
cross section is known to approach the free one.
 Above 1~GeV,  $U_S$ and $U_V$ are not known except that they are
no longer momentum independent and that they should decrease to zero.
In this case, Pauli blocking gives the only relevant contribution of the
medium and the dispersion effects are negligible.
 Therefore $U_S$ and $U_V$ will be set to zero in the calculations
of Sec.~\ref{sec:resE97-006}.

When the effective mass is accounted for, the momentum of a
nucleon participating in the rescattering process is related to
its energy $E(p)$ by Eq.~(\ref{eq:Edirac}).
 In general, ${\cal M}(s,t,u)$  is off shell. However, following
the assumption that the interaction is not appreciably modified by the
in-medium effects we use the on shell values extracted from the
vacuum pN cross section, Eq.~(\ref{eq:free-dSig}).
 In vacuum, this depends on only
two invariants that were chosen to be
$s=(p_a^\mu+h^\mu)^2$ and $t=(p_a^\mu-p_f^\mu)^2$ and computed accounting for
the dispersion relation~(\ref{eq:Edirac}).
 The energy denominators appearing in Eq.~(\ref{eq:P_dpf}) were also taken to
be equal to the total energy of the nucleon, Eq.~(\ref{eq:Edirac}). We note
that this differs from the normalization of a
Dirac spinor in the medium~\cite{waleckaANP}. However, this prescription
is consistent with the choice of using the free scattering amplitude
${\cal M}(s,t,u)$ since it provides the right normalization
in the non relativistic limit.
Finally, the in medium scattering rate is given by
\begin{equation}
  { d^3 ~ \sigma_{a N'}
    \over
    dE_f \; d \Omega_{\hat{p}_f} } =
       {1            \over      \rho_{N'} \;  v_g }
     { d p_f \over d E_f}
     { d^3 P_{aN'}   \over dp_f \; d\Omega_{\hat{p}_f} }
    \; ,
 \label{eq:SaN_dpf}
\end{equation}
where $\rho_{N'}$ is the density of the hit nucleon and
$v_g=dE_a(p_a)/dp_a=p_a/\sqrt{p_a^2 + m_{Da}^2}$ is
the group velocity of the incoming one. 
Note that the dispersion effects modify the rescattering rate,
Eq.~(\ref{eq:SaN_dpf}), in three different ways.
First, both $v_g$ and the Jacobian $dp_f /dE_f$ depend on $U_S$ and $U_V$.
Second, the density of final states for the scattered nucleons is modified
by using Eq.~(\ref{eq:Edirac})
in the energy delta function
$\delta(E_a(p_a) + E_{N'}(h) - E_f(p_f) - E_{N''}(l))$.
Third, the energies of nucleons just above $k_F$ are lowered by 
Eq.~(\ref{eq:Edirac}), which allows scattering
at energies that would otherwise be Pauli forbidden.

\begin{figure}[!t]
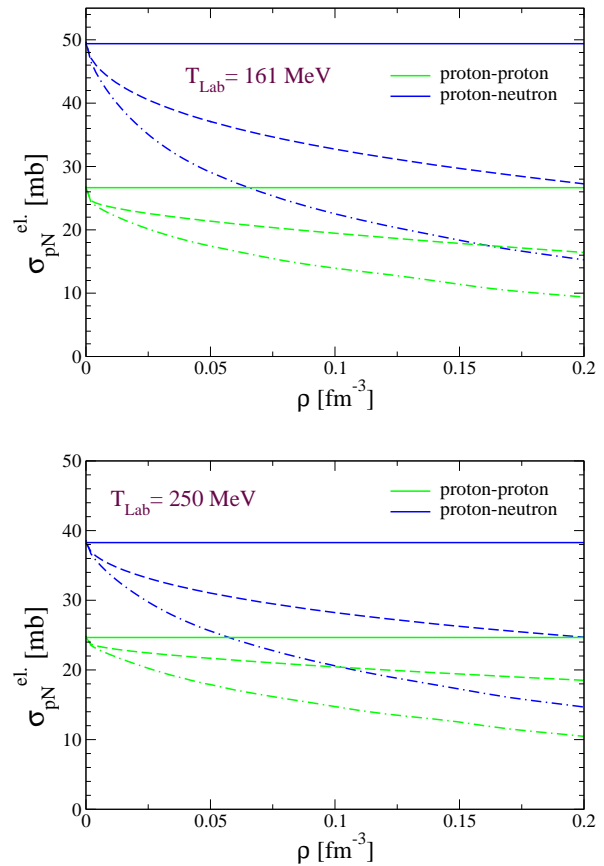

\vspace{.2in}
  \begin{center}
    \includegraphics[width=0.90\linewidth]{Plot_WM-of-Aug16_Elab=161MeV_ALL.eps}
    \vskip .5cm
    \includegraphics[width=0.90\linewidth]{Plot_WM-of-Aug16_Elab=250MeV_ALL.eps}
    \caption{ \label{fig:Stot_rho}
       (Color online) 
      In medium elastic pN cross sections as a function of
     density for laboratory energies of
     161~MeV (top panel) and 250 MeV~(bottom panel).
     The curves show the vacuum cross section (full line), the results
     obtained by accounting for Fermi spreading and Pauli blocking
     (dashed line) and by adding the interaction with the medium
     through Eq.~(\ref{eq:Edirac})~(dot-dashed line).
     }
  \end{center}
\end{figure}

\begin{figure}[!t]
\vspace{.2in}
  \begin{center}
    \includegraphics[width=0.90\linewidth]{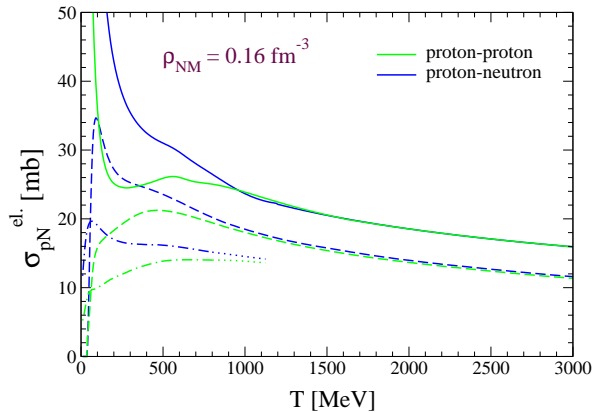}
    \caption{ \label{fig:Stot_E}
       (Color online) 
       In medium elastic pN cross sections computed for normal nuclear
     matter density as a function of the  nucleon energy.
     The curves show the vacuum cross section (full line), 
     the results obtained by including
     Pauli blocking and Fermi spreading (dashed line) and
     also the dispersion effects~(dot-dashed line).
     }
  \end{center}
\end{figure}

Figure~\ref{fig:Stot_rho} shows the in medium effective cross section 
as a function
of  density obtained by integrating Eq.~(\ref{eq:SaN_dpf})
over angle and energy.
 Two values of the energy of the incoming nucleon are considered.
 The solid line gives the vacuum cross section while the dashed line
includes the effects of Pauli blocking and Fermi spreading.
 A further reduction is produced by accounting for the dispersion effects.
 While the calculation of the Pauli blocking effects is equivalent
to the work of Ref.~\cite{VjPi},
the full cross section obtained here is somewhat smaller than
the one obtained in the corresponding non relativistic result.
A difference between the two approaches should be expected since the
present calculation is based on the Dirac effective mass, given 
by Eq.~(\ref{eq:Mdirac}). This is different form the non relativistic
definition of effective mass,
 which is instead related to the vector potential $U_V$~\cite{JamMa89}.
The effects of Pauli blocking at higher energies can be seen in
Fig.~\ref{fig:Stot_E} where the elastic cross section
is computed at normal nuclear matter density $\rho_{NM}=0.16$~fm$^{-3}$
for energies up to 3~GeV.
As it can be seen, the effects of Pauli blocking remain relevant
at large energies where they  produce a constant reduction of the
cross section.
The further reduction due to the effective mass affects the
results at low energies and tends to become
less important at 1~GeV even for constant values of $U_S$ and $U_V$.
However, it is meaningless to extend the calculation of the effective
mass effects above this energy
since the values of $U_S$ and $U_V$ are unknown in this region.

\section{Results}
\label{sec:results}

\begin{figure}[!t]
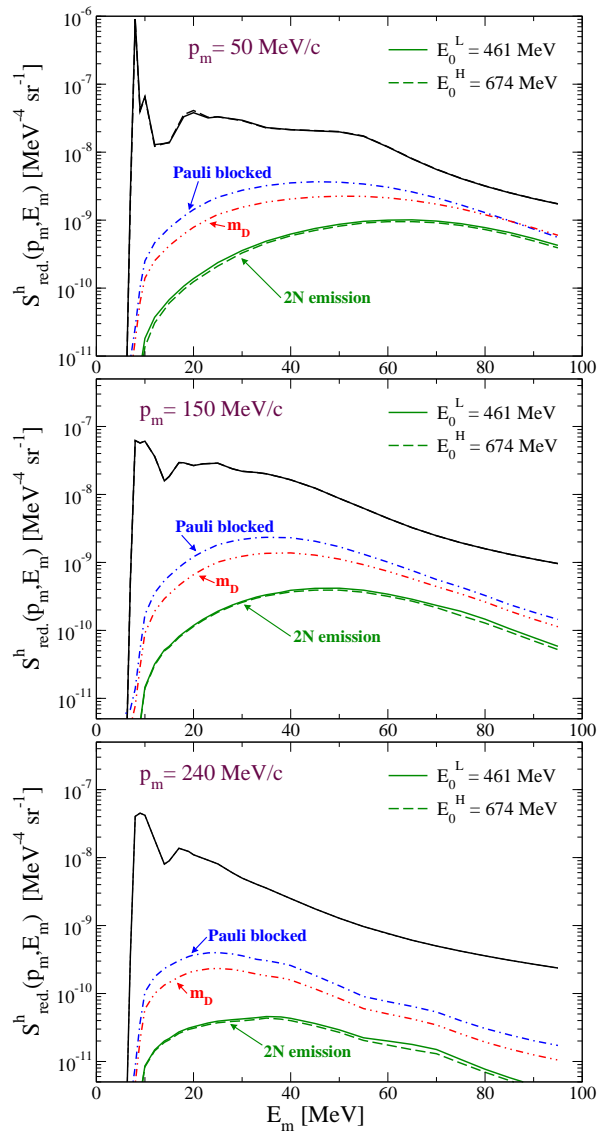

\vspace{.15in}
  \begin{center}
    \includegraphics[width=0.90\linewidth]{Plot_paper_Pb208_pm=50.eps}
    \includegraphics[width=0.90\linewidth]{Plot_paper_Pb208_pm=150.eps}
    \includegraphics[width=0.90\linewidth]{Plot_paper_Pb208_pm=240.eps}
    \caption{ \label{fig:Pb208_res}
       (Color online) 
       Theoretical results for the rescattering contribution to the reduced
      spectral strength of ${}^{208}{\rm Pb}$ for the kinematics of
      Ref.~\cite{Marcel}. The full (dashed) lines refer to the kinematics
      with lower ($E_o^L$) and higher ($E_o^H$) energy beams.
      The black lines shows the input spectral function, Eq.~(\ref{eq:Sh_MF}).
      The results obtained for rescattering with two nucleons
      emitted in the continuum are given by the green curves.
       For the kinematics $E_o^L$, the plots show the effects due 
      only to Pauli blocking and Fermi spreading (dot-dashed lines).
      If the in medium dispersion was accounted for but no two nucleon
      emission was imposed,  Eq.~(\ref{eq:P_dpf}) would give
      to the dot-dot-dashed curves.
}
\end{center}
\end{figure}

 Eq.~(\ref{fig:TotRes}) requires the knowledge of the undistorted spectral
function $S^h(p_m,E_m)$ of the target nucleus.
 For the present purposes, the strength in the mean field region
can be described as a sum over the nuclear orbitals
\begin{equation}
S^h_{MF}(p_m,E_m) = \sum_i  Z_i \; f_i(E_m) \; \left| \Phi_i(p_m) \right|^2
  \;  ,
\label{eq:Sh_MF}  
\end{equation}
where $\Phi_i(p_m)$ are the single particle wave functions, $Z_i$ are their
occupation numbers and each orbital is spread in energy according to
a Lorentzian distribution with a variable width~\cite{bror81},
\begin{eqnarray}
 f_i(E_m) &=& \frac{1}{2\pi}
  \frac{\Gamma(E_m)}{(E_m -\varepsilon_i)^2 \; + \; [\Gamma(E_m)/2]^2 }
  \; ,
\label{eq:Lorentzian}  
\\
  \Gamma(E_m) &=& \frac{a \; (E_m - E_F)^2 }{b + (E_m - E_F)^2}
  \; ,
\label{eq:Gamma_MF}  
\end{eqnarray}
where the Fermi energy was taken to be $E_F=6$~MeV,
$a=24$~MeV and $b=500$~MeV$^2$.
 The neutron spectral function was also obtained from Eq.~(\ref{eq:Sh_MF})
by including all the neutron orbits occupied in the shell structure.
 For Pb and Au no experimental data are available for neutrons,
thus the occupation numbers $Z_i$ were assigned by extrapolating
the trend measured for protons in ${}^{280}$Pb~\cite{Marcel}.
For kinematics chosen to probe medium and low missing energies, 
as those discussed in Sec.~\ref{sec:respb208}, $S^h_{MF}(p_m,E_m)$
covers all experimentally  accessible strength.
In Sec.~\ref{sec:resE97-006}, this will need to be extended by including
the distribution of nucleons in the SRC correlated region.

In the following, the rescattering yield computed from Eq.~(\ref{fig:TotRes})
has been converted to a reduced spectral  function by dividing it
by $|p_f E_f|T\sigma_{eN}^{cc1}$, evaluated for the kinematics of
the direct process [according to Eq.~(\ref{eq:PWIA})].
 This gives the straightforward correction to the model spectral function
that is due only to single rescattering effects.

\subsection{Application to ${}^{208}$Pb }
\label{sec:respb208}

The studies of Ref.~\cite{Marcel}, employed two different parallel
kinematics in which the outgoing proton was emitted in the same
direction as the momentum transfer (thus ${\bf q}$ and ${\bf p}_m$
were also parallel). The central kinetic energy of the outgoing proton was kept
constant at 161~MeV ($p_f=$570~MeV/c) in both cases. The two kinematics
differ only for the energy  of the electron beam, which was taken 
to be $E_o^H=$674~MeV for the first case and $E_o^L=$461~MeV for the other.
 These choices correspond to a virtuality $Q^2$ ranging
between 0.08 and~0.22~GeV${}^2$.

In applying the model of Secs.~\ref{rescatt_formula} and~\ref{sec:in-med_xsec}
to $(e,e'p)$ reactions one has to impose the further constraint that
both nucleons are emitted in the continuum. This is trivially satisfied
for high energy nucleons, for which the interaction with the medium 
can be neglected. When the dispersion effects are included from 
Eq.~(\ref{eq:Edirac}) one has to require that $E_l(l)>0$ in 
the integrand of Eq.~(\ref{eq:P_dpf}). Processes in which the undetected
nucleon remains bound are beyond the scope of the present work and 
would require a proper quantum mechanical treatment
to include its reabsorption  vertex.
 At low energies this should be analyzed in terms of a 
proper optical potential model.

The results of Eq.~(\ref{eq:TotRes}) for the rescattering
contributions leading to two nucleons in the continuum are
plotted in Fig.~\ref{fig:Pb208_res}.
 The yield resulting from rescattering is between one and two orders
of magnitude smaller than the direct signal, except for low missing momenta
and missing energies  above 60~MeV, where it gives a correction
of about 20\%.
The rescattering effects are also found to be independent on which of the
above kinematics is chosen.
For comparison we also show the results obtained by including only
Pauli blocking and Fermi spreading and the ones obtained when the
in medium effects are included without requiring that the undetected
nucleon is in the continuum [i.e. allowing $E_l(l)<0$ in
Eq.~(\ref{eq:P_dpf})].
 Although the latter result has no direct physical relevance, it shows
that only a part of the suppression of the rescattering effects is due
to the corrections for effective mass in the medium
(`$m_D$' curves of Fig.~\ref{fig:Pb208_res}).
 The remaining reduction is a consequence of the energy required
to reach the two-nucleon emission threshold.

\subsection{Proton knock out from the SRC region}
\label{sec:resE97-006}

\begin{figure}[!t]
\vspace{.2in}
  \begin{center}
    \includegraphics[width=0.95\linewidth]
                           {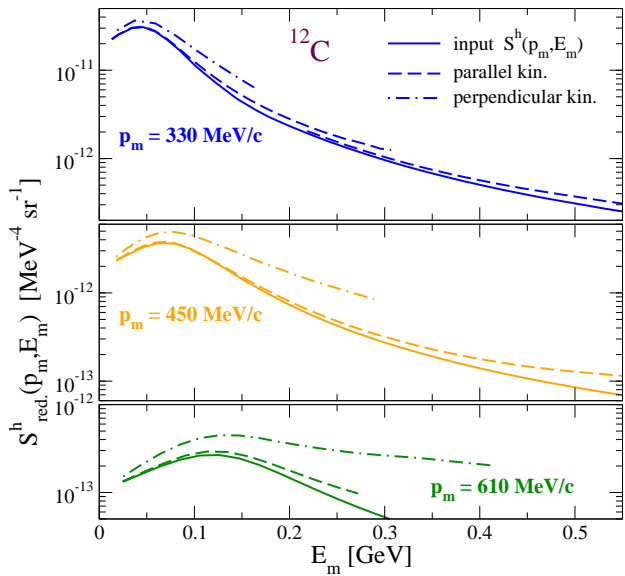}
    \caption{ \label{fig:resC12_pa_pe}
       (Color online) 
      Theoretical results for the total (direct + rescattered) reduced
     spectral strength in the
     correlated region. The results are given for parallel (dashed line) and
     perpendicular kinematics (dot-dashed line).  The full lines show
     the model spectral function, Eq.~(\ref{eq:Shtotal}), employed
     in the calculations. All panels refer to a ${}^{12}{\rm C}$ target
     and employ the same line convention.
      Note that the results for different sets of parallel kinematics
     do not always overlap exactly.
     This is mostly due to the dependence of the
     off-shell cross section $\sigma_{eN}^{cc1}$ on the
     kinematics~\cite{danielaElbaDiss}.
     }
  \end{center}
\end{figure}

\begin{figure}[!t]
\vspace{.2in}
  \begin{center}
    \includegraphics[width=0.95\linewidth]
                           {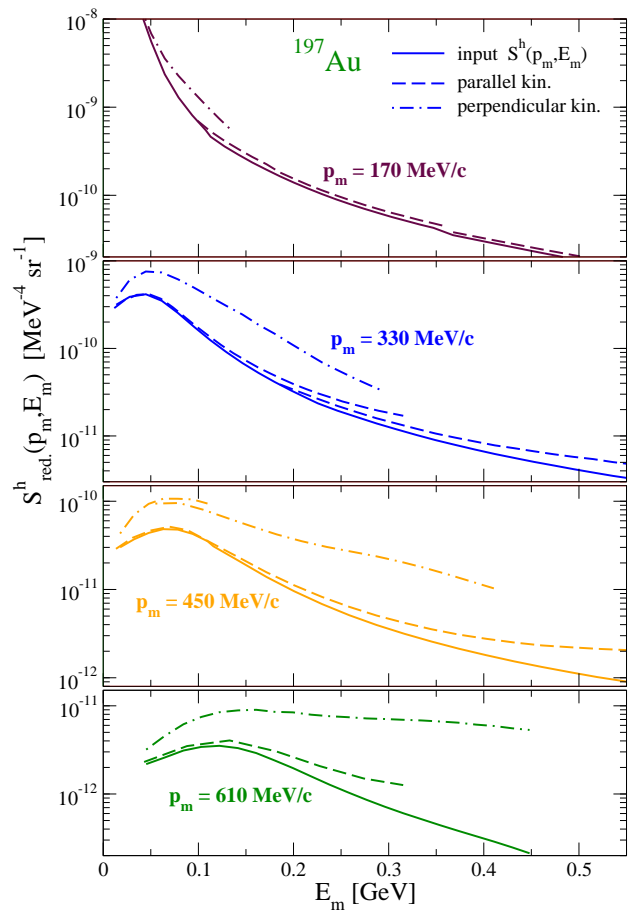}
    \caption{ \label{fig:resAu197_pa_pe}
       (Color online) 
      Theoretical results for the total reduced spectral strength in the
     correlated region. The results are given for parallel (dashed line) and
     perpendicular kinematics (dot-dashed line).  The full lines show
     the model spectral function, Eq.~(\ref{eq:Shtotal}), employed
     in the calculations. All panels refer to a ${}^{197}{\rm Au}$ target
     and employ the same line convention.
      Note that the results for different sets of kinematics do not
     always overlap exactly. This is mostly due to the dependence of the
     off-shell cross section $\sigma_{eN}^{cc1}$ on the
     kinematics~\cite{danielaElbaDiss}.
     }
  \end{center}
\end{figure}

 This section considers the results for the kinematics
of Ref.~\cite{danielaPRL}, where the aim is to directly probe SRC.
 In this case it is convenient to write the spectral function as the sum
of a mean field and a correlated part, 
\begin{equation}
 S^h(p_m,E_m) = S^h_{MF}(p_m,E_m) + S^h_{corr}(p_m,E_m) \; ,
 \label{eq:Shtotal}
\end{equation}
where $S^h_{corr}(p_m,E_m)$ describes the short-range correlated
tail at very high missing energies and momenta~\cite{WHA,Benhar}. 
 In the present work this was parametrized as
\begin{equation}
 S^h_{corr}(p_m,E_m) = 
   { C \;  e^{- \alpha \, p_m} \over [E_m - e(p_m)]^2 + [\Gamma(p_m)/2]^2}
 \label{eq:Shcorr}
\end{equation}
where $e(p_m)$ and $\Gamma(p_m)$ are smooth functions of the missing 
momentum that were chosen to give an appropriate fit to the available
${}^{12}{\rm C}(e,e'p)$ data in parallel kinematics~\cite{Frick04}.
The solid line in Fig.~\ref{fig:resC12_pa_pe} shows the model spectral function,
Eq.~(\ref{eq:Shtotal}), employed in the present calculations
for that part of the  $p_m$--$E_m$ plane where $S^h_{corr}$ dominates.
The calculation with a ${}^{197}{\rm Au}$ target employed
the same $S^h_{corr}$ of
Eq.~(\ref{eq:Shcorr}) multiplied by~$79/6$ or~$118/6$ to account for the
correct number of protons and neutrons, respectively.
This is shown for protons in Fig.~\ref{fig:resAu197_pa_pe}.
At energies close to the Fermi level the hole spectral function 
is dominated by its mean field component.
For ${}^{12}{\rm C}$ these are orbitals
in the $s$ and $p$ shells, which are known
experimentally and represent about 60~\% of the total
strength. The spectroscopic factors and wave functions
used in Eqs.~(\ref{eq:Sh_MF})--(\ref{eq:Gamma_MF}) are the ones
extracted from the world data in Ref.~\cite{LoukC12}.
Since no direct data are available for 
gold we choose to employ
the spectral function discussed above for the neighbor nucleus
${}^{208}{\rm Pb}$~\cite{Marcel} but modifying the occupation of the
last orbitals to account for the different number of nucleons in those
shells.

We have performed calculations of the rescattering contributions by
employing same sets of three parallel and two perpendicular kinematics
used in Ref.~\cite{danielaElbaDiss}.
 In the parallel case, the initial momentum of the proton ($-{\bf p}_m$)
was centered at different angles
with respect to the momentum transfered by the electron,
$\vartheta_{qpi} \sim$~25, 21~and 36~deg,
while the corresponding energies of the final proton were
centered at $E_f \sim$~1.6, 1.8~and 1.9~GeV.
This implied angles $\vartheta_{qf} \sim$~0.5, 1.5~and 10~deg, respectively,
between the final proton and the momentum transfered.
 For the two perpendicular kinematics,
the same angle $\vartheta_{qpi}\sim$~90~deg was used, while
$E_f\sim$~1.25 and 1.35~GeV  and $\vartheta_{qf} \sim$~25 and 29~deg.
The four momentum transfered by the electron was
always in the range $Q^2 \sim$~0.3--0.4~GeV$^2$. 
 More details on these kinematics are discussed in Ref.~\cite{danielaElbaDiss}.

Due to the loss of energy of the ejected nucleon at the
rescattering vertex, the spectral strength is always shifted 
toward higher missing energies.
 This is clearly visible in the results for both ${}^{12}{\rm C}$
and ${}^{197}{\rm Au}$, which are shown in Figs.~\ref{fig:resC12_pa_pe}
and~\ref{fig:resAu197_pa_pe} (the sum of direct plus rescattering
signals is plotted).
 The contribution to parallel kinematics is negligible at missing
energies below the peak of the correlated tail but it tends to become
more important for $E_m >$150--200~MeV. This confirms the expected trend
that a part of the strength seen in this region is dragged from places
where the hole spectral function is larger~\cite{E97proposal}.
 The same behavior is seen in perpendicular kinematics where, however,
rescattering effects are already relevant at small missing energies.
In this situation the direct process accounts for only 30--50\% of the
total yield obtained at the top of the correlated peak.
 At higher energies, the rescattering can overwhelm the PWIA signal
by more than an order of magnitude.
 It should be noted that for both parallel and perpendicular kinematics
the FSI become more important as the mass number increases.
In general, this is due to the average distance that the outgoing nucleon
has to travel  inside the nucleus. Thus it carries a dependence on
the nuclear radius.
\begin{figure}[!t]
\vspace{.15in}
  \begin{center}
    \includegraphics[width=0.95\linewidth]
                           {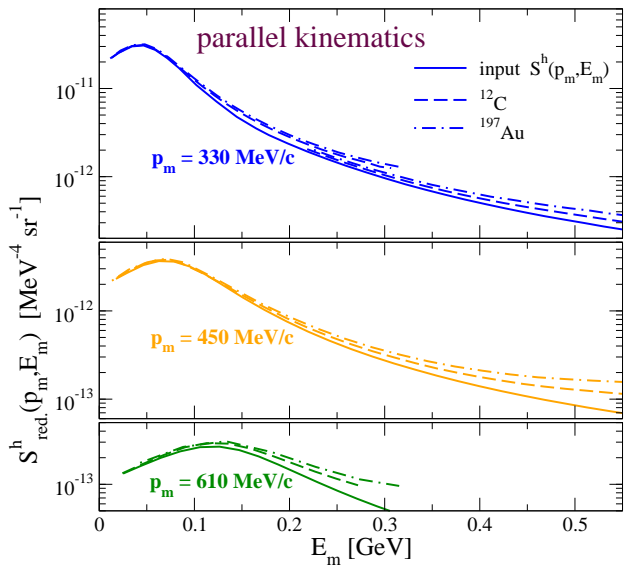}
    \caption{ \label{fig:resC_vs_Au_pa}
       (Color online) 
       Theoretical results for the total reduced spectral strength of
      ${}^{12}{\rm C}$  obtained in parallel kinematics (dashed line)
      compared to the analogous results for ${}^{197}{\rm Au}$
      (dot-dashed line), normalized to the number of protons of carbon.
       The full line shows the input spectral
      function of Eq.~(\ref{eq:Shtotal}) employed in the calculations.
       All panels employ the same line convention.
}
\end{center}
\end{figure}
\begin{figure}[!]
\vspace{.15in}
  \begin{center}
    \includegraphics[width=0.95\linewidth]
                           {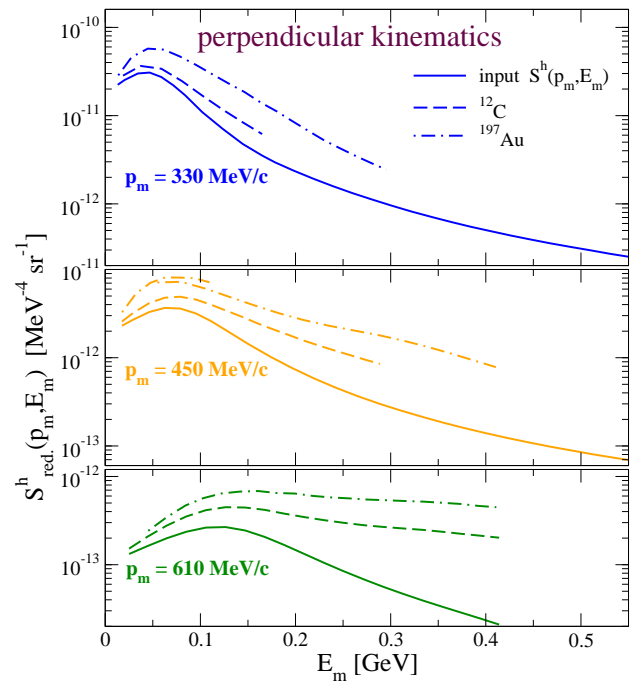}
    \caption{ \label{fig:resC_vs_Au_pe}
       (Color online) 
       Theoretical results for the total reduced spectral strength of
      ${}^{12}{\rm C}$  obtained in perpendicular kinematics (dashed line)
      compared to the analogous results for ${}^{197}{\rm Au}$
      (dot-dashed line), normalized to the number of protons of carbon.
       The full line shows the input spectral function
      of Eq.~(\ref{eq:Shtotal}) employed in the calculations.
       All panels employ the same line convention.
}
\end{center}
\end{figure}
 Figure~\ref{fig:resC_vs_Au_pa} compares the
results for both target nuclei in parallel kinematics.
For comparison the yield
for ${}^{197}{\rm Au}$ has been normalized to the same number of
protons of ${}^{12}{\rm C}$. The reduced spectral function
of gold is indeed always larger.
The same consideration applies for perpendicular kinematics
in Fig.~\ref{fig:resC_vs_Au_pe}.

\begin{figure}[t]
\vspace{.2in}
  \begin{center}
    \includegraphics[width=0.95\linewidth]
                           {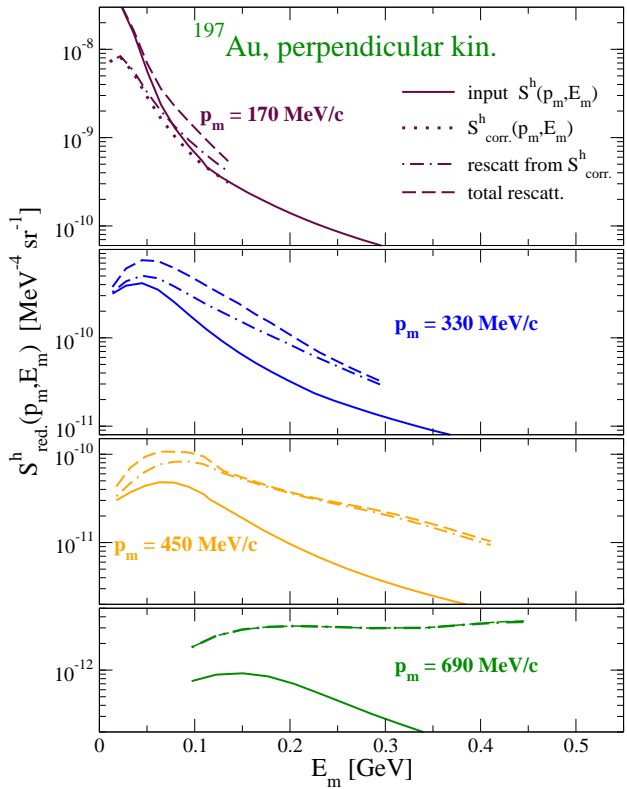}
    \caption{ \label{fig:resAu_orig_res}
       (Color online) 
       Reduced spectral strength of ${}^{197}{\rm Au}$ computed in
      perpendicular kinematics as generated by the full spectral function of
      Eq.~(\ref{eq:Shtotal}) (dashed line) or by the sole correlated
      part $S^h_{corr}$ (dot-dashed line).
       The full line shows the model spectral function
      of Eq.~(\ref{eq:Shtotal}). Its mean field component, $S^h_{MF}$,
      is not visible in this plot except for small missing momenta.
       Note that the dashed and dot-dashed lines overlap in the bottom panel.
       All panels employ the same line convention.
}
\end{center}
\end{figure}

To study the origin of the rescattered strength,
the calculations were repeated for gold by neglecting the mean field
orbitals, $S^h_{MF}$, in Eq.~(\ref{eq:Shtotal}).
For this nucleus the mean field strength extends to large missing energies,
up to $\sim$100~MeV. Thus rescattering effects can easily spread it
into the correlated region.
Figure~\ref{fig:resAu_orig_res} compares the theoretical reduced spectral
strength of Au with the analogous result obtained
when $S^h_{MF}$ is included.
 As one can see, relevant contributions from the mean field appear
for momenta up to about 500~MeV/c.
 The results at higher missing momenta are completely dominated by
the rescattering from the correlated tail $S^h_{corr}$ into the
correlated region itself.
The situation is instead different in parallel kinematics where the
present results for two-step rescattering do not shift any strength
from the mean field region, even for a large nucleus like Au.
 Such large shift appears to be energetically forbidden do to the 
large energy of the scattered proton adopted in these kinematics.

\section{Conclusions}
\label{sec:concl}

A proper understanding of the relevance of FSI and their dependence on
the kinematics is important in modern  $(e,e'p)$ experiments that attempt
to observe the correlated strength at medium and high missing energies.
 The present work suggests a semiclassical approach to compute the 
effects of two-step rescattering, which is one of the leading contributions
at high proton energies, and applies it to investigate its consequences
for the kinematics of two different experiments.

The model assumes a PWIA for the electromagnetic vertex, in which the
struck nucleon is described by the full hole spectral distribution
of the target nucleus.
 This gives the possibility of investigating how FSI shift the original
strength from the direct process within the missing energy and momentum
plane.
 The absorption effects of the medium were accounted for
by means of transparency factors. The rescattering is described in
terms of the differential nucleon-nucleon cross section modified in order
to account for Pauli blocking and Fermi spreading effects. The dispersion
effects due to the nuclear medium have been included from DBHF 
results at the energies where they are relevant.

For kinematics involving outgoing protons of the order of few hundreds
of MeV, the
present model was employed to estimate the reappearance of strength
through inelastic channels that lead to two-nucleon emission.
 In the reaction $^{208}$Pb$(e,e'p)$ the overall
effects were seen to be more than an order of magnitude smaller than the
direct signal, slightly increasing for small missing momenta and
missing energies above 60~MeV.
 This supports an analysis of the experimental data based on usual
distorted wave calculations.

The same model was applied for the kinematics of the E97-006 experiment
at JLab that focussed on the SRC distributions at high momenta.
Calculations were performed for ${}^{12}{\rm C}$
and ${}^{197}{\rm Au}$ targets, which have different radii.
In general, rescattering was found to be much smaller in parallel
kinematics than in perpendicular ones.
In the latter case a large amount of strength is shifted from regions
where the spectral function is big to regions where it is smaller, thus
overwhelming the experimental yield from the direct process.
This confirms the studies of Ref.~\cite{E97proposal}.
The contribution from rescattering effects is also seen to increase with
the nuclear radius.
The rescattering of nucleons originally emitted form the
mean field orbitals was found to be important  in perpendicular
kinematics and for missing momenta lower than $\sim$500~MeV/c.
 No such a large shift of strength was found in the parallel case.
 The remaining  effects of rescattering are due to 
a rearrangement of the spectral strength within the correlated 
tail itself.

The present results provide a good first insight in the redistribution
of strength due to FSI in $(e,e'p)$ reactions.
However, it is clear that in order to properly explain the real experimental
yield observed at high missing energies and for heavy nuclei
other effects beyond the two-step rescattering need
to be addressed~\cite{NicolaevGL,Ciofi4He}.
Relevant extensions of the present formalism would include corrections
from meson exchange and nucleon excitation processes, as well as
an investigation of the importance of multiple rescattering for heavy
nuclei in parallel kinematics. Work in this direction is in
progress.

~ \\ 

\acknowledgments
We would like to acknowledge several useful discussions with
V.~R.~Pandharipande, D.~Rohe and I.~Sick, and thank W.~H.~Dickhoff
for comments on a preliminary version of the manuscript.
This work is supported by the Natural
Sciences and Engineering Research Council of Canada (NSERC) and
by the ``Stichting voor Fundamenteel Onderzoek der Materie (FOM)'',
which is financially supported by the ``Nederlandse Organisatie voor
Wetenschappelijk Onderzoek (NWO)''.


\end{document}